\documentclass[doublecol]{epl2} 
\newcommand{\be}{\begin{equation}}
\newcommand{\ee}{\end{equation}}
\newcommand{\bea}{\begin{eqnarray}}
\newcommand{\eea}{\end{eqnarray}}
\newcommand{\hf}{\frac12}
\newcommand{\nn}{\nonumber\\}
\def\journal#1#2#3#4{\REVIEW{#1}{#2}{#4}{#3}}
\def\mr#1{{\mathrm{#1}}}
\def\v#1{{\bf{#1}}}
\def\eq#1{(\ref{#1})}

\def\la{\langle}
\def\ra{\rangle}
\def\ab{\bar a}
\def\bk{k_\mr{gas}}
\def\ha{\hat a}

\def\hsid{\hat\psi^\dagger}

\def\hsi{\hat\psi}
\def\hsib{\hat{\psi^\dagger}}
\def\hG{{\hat G}}
\def\hJ{{\hat J}}
\def\ih{\frac{i}{\hbar}}
\def\psid{\psi^\dagger}
\def\operc{{\cal C}}

\def\tG{\tilde G}
\def\Tr{\mathrm{Tr}}

\title{Decoherence and damping in ideal gases}
\shorttitle{Ideal gases} %Insert here a short version of the title if it exceeds 70 characters

\author{Janos Polonyi}
\shortauthor{J. Polonyi}

\institute{Strasbourg University, High Energy Theory Group, CNRS-IPHC,
23 rue du Loess, BP28 67037 Strasbourg Cedex 2 France
}

\abstract{
The particle and current densities are shown to display damping and undergo
decoherence in ideal quantum gases. The damping is read off from the equations
of motion reminiscent of the Navier-Stokes equations and shows some formal similarity
with Landau damping. The decoherence leads to consistent density and current
histories with characteristic length and time scales given by the ideal gas.
}

\begin{document}

\maketitle

\section{Introduction}
Realistic, complex systems usually display separate length scales. In fact,
a shorter length scale characterizes the elementary excitations, e.g.
the average separation of particles. The interactions generates 
another, longer length scale, such as the mean free path. These scales
are separated by some power of the coupling constants and the measurements, 
carried out by macroscopic devices, uncover the manifestations of the longest 
length scale, c.f. collective phenomena. 

One usually thinks of the shortest,
microscopic scale as the reflection of the dynamics of some trivial, 
non-interacting particle system. It is shown in this letter that contrary
to this view the observables used to diagnose the system may create strong, 
order one effective interaction vertices for undistinguishable, noninteracting 
particles. What happens is that the correlations generated by the exchange 
statistics of the these particles produce involved correlations for local 
observables and create the illusion of some interactions.

The case of a quantum ideal gas is consedered in this work where the effective
dynamics of the particle and current densities is sought in the absence of any
interaction. It is well known that for non-interacting particles these operators have one-loop, 
connected Green functions in arbitrary high order in contrast to
the elementary fields whose higher order Green functions factorize according
to Wick's theorem. The non-factorization of the Green functions indicates entanglement
and suggests that the effective dynamics of the particle and the current density as
local degrees of freedom might be highly nontrivial. The main result of
this work is that decoherence and damping arise in the effective dynamics of 
the particle and current density, inherent in an ideal quantum gas.

\section{Model and method}
Let us cope first with the issue of interactions in a free system, described by a coordinate 
$x$ and the action $S[x]$. Suppose that we are interested in the dynamics of a collective 
coordinate $y=F(x)$. It is easy to see that the extended action
\be
S[x]\to S[x,y]=S[x]-K\int dt[y(t)-F(x(t))]^2
\ee
with $K$ being a constant generates the correct dynamics of both coordinates. Note that the second 
term in the right hand side contains higher than quadratic powers of $x$ if the collective 
coordinate $y$ is a non-linear function of $x$. The action $S[x,y]$ offers another
advantage, it allows us to consider the free coordinate $x$ as the environment of the
collective coordinate $y$.

We shall consider in this work a model of non-interacting particles correlated only by quantum 
statistics and investigate the dynamics of the particle density and current density, constituting
a conserved 4-vector $j^\mu=(n,\v{j})$, viewed as collective coordinates (subsystem) 
related to the particle degrees of freedom (environment) in a non-linear way, by means 
of the open time path (OTP) formalism. This is a slightly extended version of Schwinger's
close time path (CTP) method \cite{schw} which provides the required general 
framework for exploring the open subsystem dynamics. By opening the closed time
path at the final time we gain access to the density matrix.

The CTP formalism has alread been used to establish decoherence and 
the consistency of histories \cite{calzettahua} and to derive the transport 
equations \cite{calzettahu}. While these approaches aim at the collective mode 
dynamics arising from genuine interactions our goal is the construction of 
the effective dynamics of composite operators, particle density and current, 
in ideal gases. A further simplification is that the 1PI formalism is used below
when the experimentally privileged local observables, the particle density and 
current operators are considered as the first few terms contributing to the 
McLaurin series of the two point functions in the Fourier space. 
We can make in this manner a short cut and 
consider the effective dynamics of the local density and current operators
instead of treating the whole complexity of the two-point functions in the 
2PI formalism. The result is that we do not have to follow the traditional,
phenomenologically motivated way of obtaining the hydrodynamical equations 
from the energy-momentum conservation but can approach the problem in a more
natural manner by considering directly the variational equations of the 
effective action for the density and current.

The path integral representation of the CTP generator functional for the connected Green 
functions of the 4-current density  $j^\mu=\psid\operc^\mu_x\psi$ \cite{ed} reads
\be\label{barectppi}
e^{\ih W[\ha]}=\int D[\hsi]D[\hsib]e^{\ih\sum_{\sigma\sigma'}
\hsib^\sigma(\hG^{-1\sigma\sigma'}+\delta^{\sigma\sigma'}a^\sigma\operc)\hsi^{\sigma'}},
\ee
where $\hsi$ and $\ha$ are CTP doublets and $\hG^{-1}$ denotes the inverse CTP 
propagator. We use the condensed notation where the functions defined in space-time
and operators acting on them are considered as vectors and matrices, respectively.
The integration over the fields $\hsi$ and $\hsid$ yields the non-polynomial influence functional
\be
W[\ha]=i\xi\hbar\Tr\ln(\hG^{-1}+\ha\operc).
\ee
The resulting connected Green functions with an arbitrary large number of external legs
correspond to the one-loop structures build up by a single particle line, mentioned 
in the Introduction. We assume that the fluctuations of the current are small and
retain the quadratic term
\be\label{quadrw}
W^{(2)}[\ha]=\hJ_\mr{gr}\ha-\hf\ha\tG\ha,
\ee
where $J^{\sigma\mu}_\mr{gr}=i\xi\hbar\Tr[\hG\operc^{\sigma\mu}_x]=(\sigma n_0,\v{0})$,
$n_0$ being the particle density in the rest frame. The particle-hole propagator
\be\label{prtchpr}
\tG^{(\sigma\mu)(\sigma'\mu')}_{xx'}
=i\xi\hbar\Tr[\hG^{\sigma'\sigma}\operc^\mu_x\hG^{\sigma\sigma'}\operc^{\mu'}_{x'}].
\ee
contains the exchange statistics factor $\xi$. Exploiting translational and rotational invariance 
one finally ends up with the CTP structure
\be\label{tgres}
\tG=\pmatrix{L&iS^-\cr-iS^-&-L}-iS^+\pmatrix{1&1\cr1&1}.
\ee
where each block $X=L,S$ is a tensor
\be
X^{\mu\nu}_{\omega,\v{q}}=\pmatrix{X^{tt}_{\omega,q}
&\v{q}X^{ts}_{\omega,q}\cr\v{q}X^{st}_{\omega,q}&
LX^L_{\omega,q}+TX^T_{\omega,q}},
\ee
where $L=\v{\nabla}\otimes\v{\nabla}/\Delta$, $T=1-L$,
\be\label{lindh}
L^{\mu\nu}_q=n_s\hbar P\int_\v{k}n_\v{k}\left[\frac{F^{\mu\nu+}_{\v{k},\v{q}}}
{\omega-\frac{\hbar q^2}{2m}-\frac{\hbar\v{k}\v{q}}m}
-\frac{F^{\mu\nu-}_{\v{k},\v{q}}}{\omega+\frac{\hbar q^2}{2m}+\frac{\hbar\v{k}\v{q}}m}\right],
\ee
with $\int_\v{k}=\int d^3k/(2\pi)^3$,
\be
F^{\pm\mu\nu}_{\v{k},\v{q}}=\pmatrix{1&\mp\frac{\hbar}{m}\left(\v{k}+\frac{\v{q}}{2}\right)\cr
\mp\frac{\hbar}{m}\left(\v{k}+\frac{\v{q}}{2}\right)&\frac{\hbar^2}{m^2}
\left(\v{k}+\frac{\v{q}}{2}\right)\otimes\left(\v{k}+\frac{\v{q}}{2}\right)}
\ee
is the CTP generalization of the Lindhart function, $S^\pm=R^+\pm R^-$,
\be\label{imag}
R^{\pm\mu\nu}_{\omega,\v{q}}=n_s\pi\hbar\int_\v{k}\delta(\pm\omega-\omega_{\v{k}+\v{q}}+\omega_\v{k})
n_\v{k}(1+\xi n_{\v{k}+\v{q}})F^{\mu\nu\pm}_{\v{k},\v{q}},
\ee
$n_\v{k}=1/(e^{\beta\omega_\v{k}}-\xi)$, $n_s=(2s+1)$ \cite{hajdu}. 
The retarded propagator is given by $G^r=-L+iS^-$.

\section{Bare dynamics}
The bare action $S^B[\hJ]$ for the current expectation value 
$\hJ=\delta W[\ha]/\delta\ha$, defined by
\be\label{bareactpi}
e^{\ih W[\ha]}=\int D[\hJ]e^{\ih S^B[\hJ]+\ih\hJ\ha},
\ee
reads in the quadratic approximation 
\be\label{quadrbareact}
S^B[\hJ]=\hf(\hJ-\hJ_\mr{gr})\tG^{-1}(\hJ-\hJ_\mr{gr}).
\ee
The OTP formalism is realized by sending the final time of the underlying CTP 
scheme to infinity and considering this action which contains no implicit boundary 
conditions at the final time anymore for arbitrary, not necessarily closed pairs of 
trajectories $(J^+,J^-)$. Parametrizing the source as $a^\pm=a(1\pm\kappa)/2\pm\ab$ 
\cite{ed} and expressing $\hJ$ by its CTP components which couple to $a$ 
(physical current) and $\ab$ (auxiliary current) the current-source coupling 
$\ha\hJ=aJ+\ab J^a$ gives $J=(J^++J^-)/2+\kappa(J^+-J^-)/2$, $J^a=J^+-J^-$.
The inversion of the last two equations yields the form of the original 
current CTP components $J^\pm=J-(\kappa\mp1)J^a/2$ in the bare action in 
terms of the physical and the auxiliary fields $J^\mu$ and $J^{a\mu}$.
The calculation of $\tG^{-1}$ of the bare action is tedious but straightforward and one
finally obtains $S^B=\Re S^B+i\Im S^B$, 
\bea\label{bareactri}
\Re S^B&=&\hf\int_{\omega,\v{k}}\biggl[
\frac{L^{tt}_{\omega,k}n^*_{\omega,\v{k}}n^a_{\omega,\v{k}}
-iS^{tt-}_{\omega,k}n^*_{\omega,\v{k}}n^a_{\omega,\v{k}}}
{(1+z^2)[(L^{tt}_{\omega,k})^2+(S^{tt-}_{\omega,k})^2]}\nn
&&+\frac{L^T_{\omega,k}\v{j}^{T*}_{\omega,\v{k}}\v{j}^{Ta}_{\omega,\v{k}}
-iS^{T-}_{\omega,k}\v{j}^{T*}_{\omega,\v{k}}\v{j}^{Ta}_{\omega,\v{k}}}
{(L^T_{\omega,k})^2+(S^{T-}_{\omega,k})^2}\biggr],\nn
\Im S^B&=&\frac14\int_{\omega,\v{k}}\biggl[
\frac{S^{tt+}_{\omega,k}n^{a*}_{\omega,\v{k}}n^a_{\omega,\v{k}}}
{(1+z^2)[(L^{tt}_{\omega,k})^2+(S^{tt-}_{\omega,k})^2]}\nn
&&+\frac{S^{T+}_{\omega,k}\v{j}^{Ta*}_{\omega,\v{k}}\v{j}^{Ta}_{\omega,\v{k}}}
{(L^T_{\omega,k})^2+(S^{T-}_{\omega,k})^2}\biggr],
\eea
with $z=m\omega/\hbar\bk k$ where $\bk$ is the characteristic wave vector of the one-particle distribution function $n_k$, 
i.e. the Fermi wave-vector $k_F$, for fermions and the thermal wave-vector
$\sqrt{mk_BT}/\hbar$ for bosons without condensate.

\section{Consistency and decoherence}
The usual way of discovering decoherence is either to approximate 
the system-environment interations as a successive scattering process 
\cite{zeh,joos} or to eliminate a direct product factor of the Hilbert 
space in some simple model \cite{zureke,zurekk}.
The former approach offers a suggestive view of the dynamical origin
of decoherence but remains rather qualitative. The latter, more
formal method orients our interest towards the degeneracy as the
source of decoherence. But to establish a sufficiently degenerate
environment one needs a large number of degrees of freedom and such
a system can be handled by means of quantum field theory only. This is the 
point where the OTP/CTP schemes arise in a natural manner because they can 
handle the (reduced) density matrix by means of Green functions in many-body 
systems.

$\Im S^B$, which is equal to the imaginary part of the influence functional, 
controlls the consistency \cite{griffiths,omnes,gellmann} of the pairs of OTP 
trajectories of the subsystem, running  to the final time $t_f$. Furthermore, the 
dependence of $\Im S^B$ on $t_f$ characterizes the building up of the decoherence 
\cite{zeh,zureke,zurekk,joos} of the environment states $|e^\pm(t_f)\ra$, 
generated from $|e\ra$ by the time evolution in the presence of
fixed system trajectories, $\la e^+(t_f)|e^-(t_f)\ra=\exp iS^B/\hbar$.
The reduced density matrix of the subsystem 
$\la\psi^+|\Tr_e\rho(t_f)|\psi^-\ra=\exp iW[\hJ]/\hbar$ is given by Eq. \eq{barectppi}. 
If the decoherence of an environment state is robust then it is referred to as
pointer state \cite{zureke}. A necessary condition for robustness is the persistence
of the decoherence for macroscopic times, i.e a sufficiently large value of $\Im S^B$
for trajectories with $\omega\to0$. In the present consideration the subsystem is
characterized by the 4-vector $(n_{t,\v{k}},\v{j}_{t,\v{k}})$ for each fixed
wave vector $\v{k}$, and $\Im S^B$ indeed controlls the consistency of pairs of
small amplitude oscillatory trajectories.

\begin{figure}
\onefigure[scale=.5]{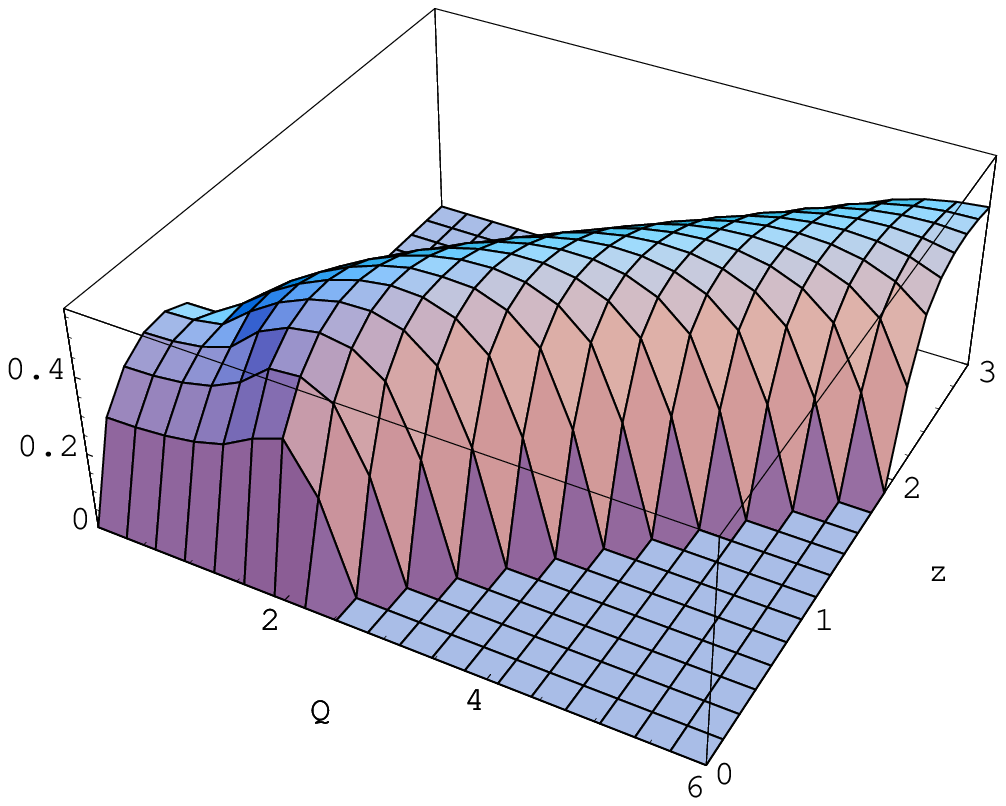}

\onefigure[scale=.5]{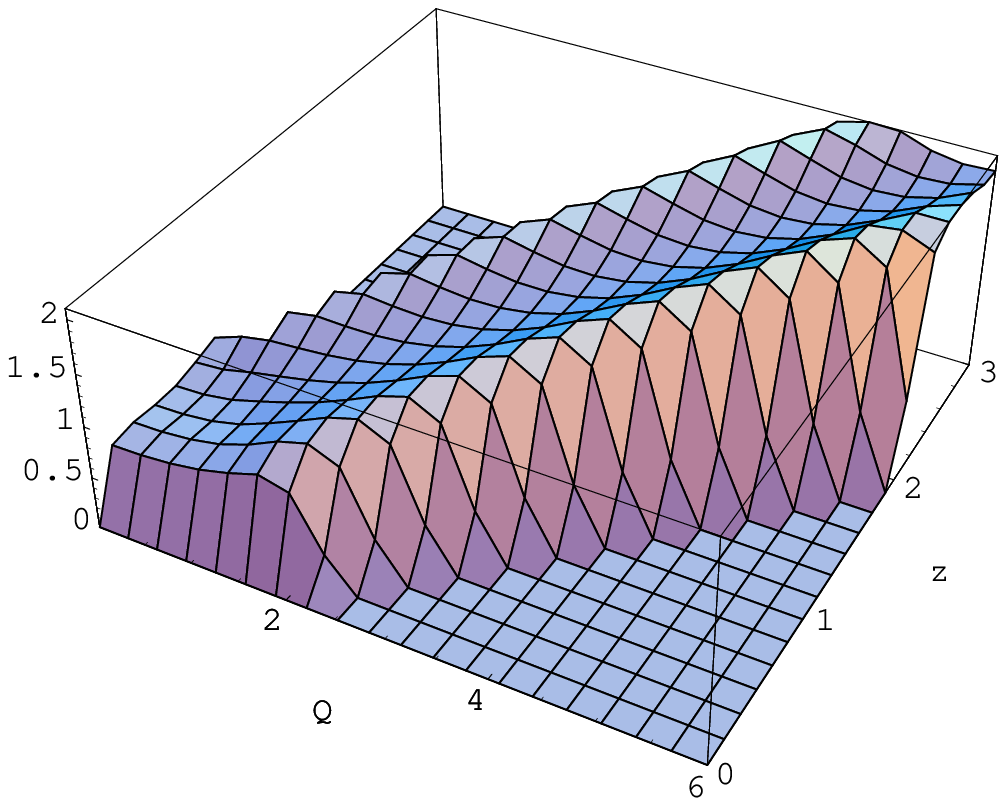}
\caption{$D^{tt}$ (upper) and $D^T$ (lower) as functions of $Q$ and $z$.}\label{dtt}
\end{figure}

To construct a measure for consistency and decoherence we choose a time dependent 
state in which the expectation value $J^\mu$ is a plane wave with amplitude 
$n_{\omega,\v{k}},\v{j}_{\omega,\v{k}}$ and consider the overlap 
with states with infinitesimally different amplitudes. For a single mode, the bare
functional integral of Eq. \eq{bareactpi} assumes the form of ordinary integrals leading 
to the expectation values $\la n^*n\ra=-2\hbar S^{tt+}(1+z^2)$, 
$\la n^*n^a\ra=2i\hbar(L^{tt}+iS^{tt-})(1+z^2)$,
$\la\v{j}^{T*}\v{j}^T\ra=-2\hbar S^{T+}$, 
$\la\v{j}^{T*}\v{j}^{Ta}\ra=2i\hbar(L^T+iS^{T-})$
(the indices $\omega$, $\v{k}$ being suppressed). The inverse of the 
second moment of $J^a$ will be chosen as a measure of consistence or 
decoherence \cite{coulomb}. Since $\la n^{a*}n^a\ra=\la\v{j}^{Ta*}\v{j}^{Ta}\ra=0$ 
in the absence of external source we use instead the ratios
\bea\label{absdec}
(D^{tt})^2&=&\frac{\hbar|\la n^*n\ra|}{|\la n^*n^a\ra|^2\ra}
=\frac{|S^{tt+}|}{2(1+z^2)[(L^{tt})^2+(S^{tt-})^2]},\nn
(D^T)^2&=&\frac{\hbar|\la\v{j}^{T*}\v{j}^T\ra|}{|\la\v{j}^{T*}\v{j}^{Ta}\ra|^2}
=\frac{|S^{T+}|}{2[(L^T)^2+(S^{T-})^2]},
\eea
for this purpose. Large $D$ indicates that the quantum fluctuations around the
trajectory with given, oscillatory plane wave expectation value for $J$ are 
distributed by mainly classical probabilities. They decohere and the neighboring 
trajectories tend to be consistent. One expects that classical probabilities
are recovered at long distances and times. The time and distance scale of the
restoration of classical probabilities can be estimated by the value of
$1/\omega$ and $1/|\v{k}|$ where these ratios become large. In the absence of 
other dimensional constants than $\hbar$, the particle mass $m$ and the density 
$n$ of the ideal gas these scales are in the order of magnitude of 
$\hbar/\epsilon_{\bk}$, $1/\bk~$ and fall into the quantum domain. Thus 
consistency or decoherence occur at microscopic scale for ideal gases and
the quantum-classical crossover remains inaccessible for macroscopic devices. 

\begin{figure}
\onefigure[scale=.5]{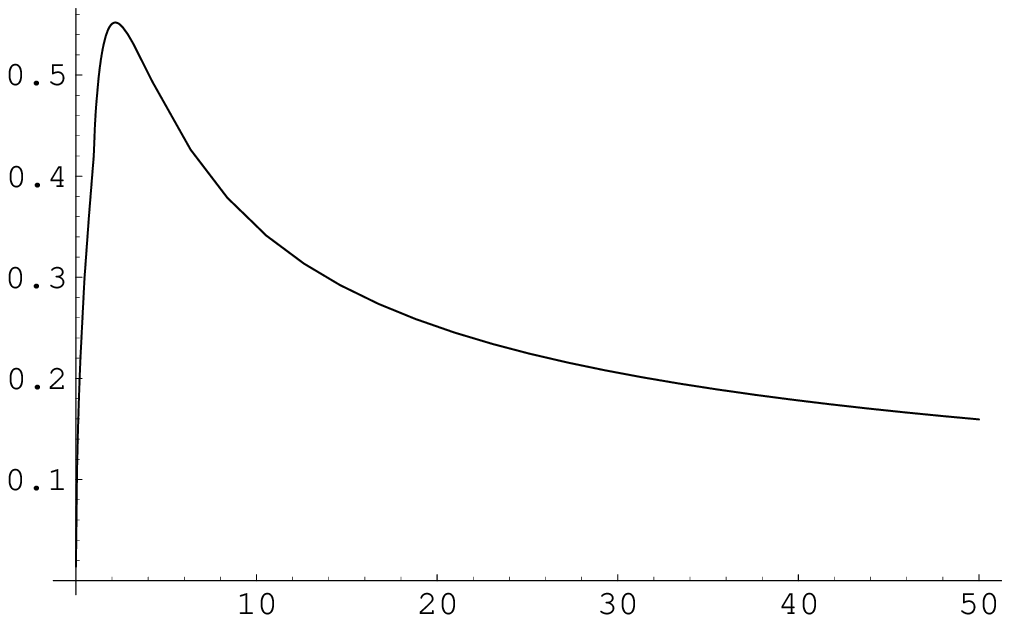}

\onefigure[scale=.5]{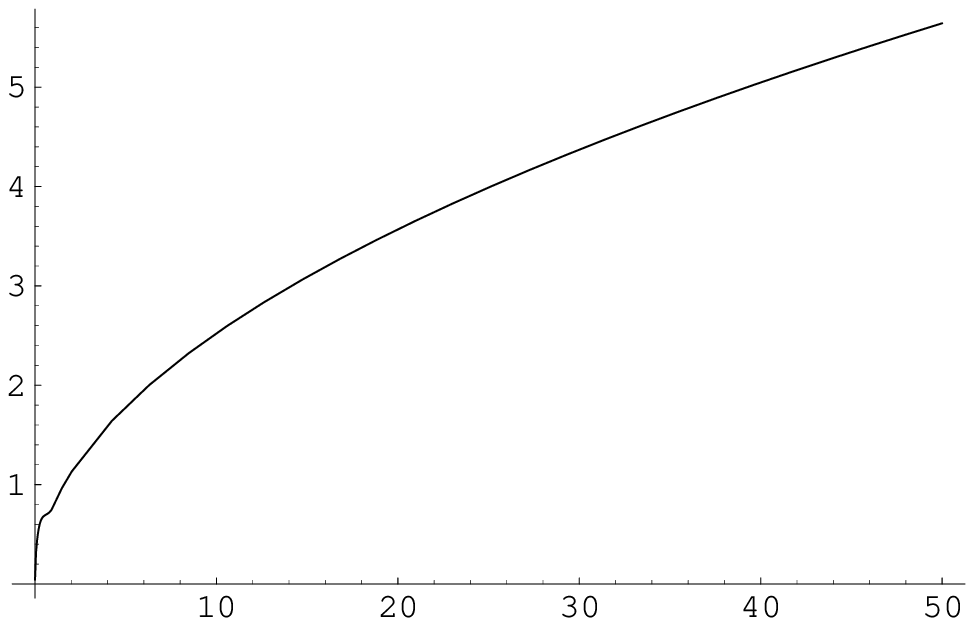}
\caption{$D^{tt}$ (upper) and $D^T$, (lower) plotted along the 
line $z=Q/2$ as functions of $Q$.}\label{dttd}
\end{figure}

The ratios given by Eqs. \eq{absdec} are plotted in Fig. \ref{dtt}. They are 
nonvanishing in two strips, starting from the origin of the plane $(Q,z)$ and 
having slopes $\pm1/2$ and width 2. The center of the strips is the straight 
line $z=\pm Q/2$, corresponding to the mass-shell condition of the environment, 
$\hbar\omega=\hbar^2q^2/2m$. The value of the functions along this line
is shown in Fig. \ref{dttd}. The most consistent density trajectories are
around $(Q,z)=(2,1)$. The classical features might be enhanced in the
IR directions by means of appropriate interactions. The current trajectories
become more consistent when we move towards the UV direction. This does not
imply classical behavior at short distances or times, the decoherence and
consistency being necessary but not sufficient conditions for classical
physics. There are few particles involved at short space or time scales and
the quantum fluctuations are important. Furthermore they do not allow the 
recovery of the deterministic classical laws. The qualitative difference of $D^{tt}$ and $D^T$
suggests that the scale of the quantum-classical crossover is determined by 
the density rather than the current fluctuations.

\section{Equations of motion}
The equation of motion for the expectation value of the 4-current expectation value
$J=\delta \Re W[a,\ab]/\delta a$ is obtained from the effective action
$\Gamma[J]=\Re W[a,\ab]-a\cdot J$ whose quadratic part in the fields is
\be
\kappa\Gamma[J]=J\tG^{n-1}(\tG^r\ab+J_\mr{gr})-\hf J\tG^{n-1}J.
\ee
Clearly, the effective action is well defined for $\kappa\not=0$ only \cite{ed}. 
The corresponding equation of motion
\be
\tG^{r-1}(J-J_\mr{gr})=\ab
\ee
is diagonal in Fourier space. The highly complicated, non-polynomial
equations simplify considerably in the hydrodynamical limit defined by
$\tilde k=k/\bk\ll1$, $z\ll1$,
\bea\label{nvms}
-g^0\ab^0&=&\left(a_0+a_z\frac{m}{\hbar\bk}\frac{i\omega}k+\frac{a_{kk}}{\bk^2}k^2\right)n,\nn
-g^0\bar{\v{a}}^L&=&\left(a_0+a_z\frac{m}{\hbar\bk}\frac{i\omega}k+\frac{a_{kk}}{\bk^2}k^2\right)\v{j}^L,\nn
-g^T\bar{\v{a}}^T&=&\left(b_0+b_z\frac{m}{\hbar\bk^2}\frac{i\omega}k+\frac{b_{kk}}{\bk^2}k^2\right)\v{j}^T,
\eea
where $g^0=n_smk_F/4\pi^2$, $g^T=g^0\hbar^2k_F^2/m^2$. For zero temperature 
fermions the numerical constants 
are $a_0=1/2$, $a_z=\pi/4$, $a_{kk}=1/24$, $b_0=3/2$, $b_z=9\pi/4$ and 
$b_{kk}=39/128$. These equations are reminiscent of the linearized
Navier-Stokes equation of hydrodynamics. The different coefficients 
appearing for the longitudinal and the transverse current
introduce different bulk and shear viscosity effects.

To see the nature of the dissipative terms appearing in Eqs. \eq{nvms} more clearly 
we consider a stationary flow in the $w$-direction which depends on the $x$-coordinate only.
For the static deviation of the density and the current density from the corresponding 
ground state expectation values Eqs. \eq{nvms} yield
\bea\label{stateqs}
n_\v{k}&=&g^0\left[\left(\frac1{\tilde k}-\frac{\tilde k}4\right)
\ln\left|\frac{2-\tilde k}{2+\tilde k}\right|-1\right]a^0_\v{k},\\
\v{j}^T_\v{k}&=&g^T\left[\frac{\tilde k^2}{16}+\frac1{\tilde k}
\left(1-\frac{\tilde k^2}4\right)^2\ln\left|\frac{2-\tilde k}{2+\tilde k}\right|-\frac5{12}\right]\bar{\v{a}}^T_\v{k}\nonumber
\eea
The longitudinal component of the current, $\v{j}^L$, can be reconstructed from these equations
by means of the continuity equation for the current. Note that Eqs. \eq{stateqs} contain all 
higher order derivatives. The only approximation involved is the linearization of the equation 
of motion. For the sake of simplicity \eq{stateqs} is approximated by
\bea\label{stateqsa}
n_\v{k}&\approx&-2g^0e^{-\frac{Q^2}6}\ab^0_\v{k},\nn
\v{j}^T_\v{k}&\approx&-\v{z}g^T\frac23e^{-\frac{Q^2}3}\ab^T_\v{k}.
\eea
We choose a weak external source
\be\label{infsource}
\ab^X_\v{x}=\frac{u^X}{(2\pi\ell_\mr{ext}^2)^{3/2}}e^{-\frac{x^2}{2\ell_\mr{ext}^2}},
\ee
where the index $X$ stands for $0$ or $T$. It yields a Gaussian flow,
\bea\label{homfl}
n_x&=&-2\frac{g^0u^0}{(2\pi\ell_\mr{flow}^2)^{3/2}}
e^{-\frac{x^2}{2\ell_\mr{flow}^2}},\nn
\v{j}^T_x&=&-\v{z}\frac23\frac{g^Tu^T}{(2\pi\ell_\mr{flow}^2)^{3/2}}e^{-\frac{x^2}{2\ell_\mr{flow}^2}},
\eea
with $\ell_\mr{flow}=\sqrt{\ell_\mr{ext}^2+1/3k^2_F}$. Such a spread of the
external perturbation is a collective phenomenon due to the presence of $k_F$ in its
length scale and is reminiscent of a diffusive process.

It is easy to find the dynamical origin of this diffusion, it is the well known spread of 
wave-packets of a free particle. In fact, the plane wave states not only serve as a useful
basis for free particles, they represent states which, one-by-one, 
produce robust expectation values. Any smearing of the wave function of this state  
in the momentum representation to an ordinary wave-packet which is a linear superposition of 
plane waves without some special fine tuning of the relative phases leads to the complete
loss of localization as the time passes, according to the Riemann-Lebesgue lemma. The
same spread of the individual plane wave components spreads any local observable such as the 
density or current. The external source for the three-current, $\v{a}$,
can be interpreted as a space-dependent shift of the momentum variable, eg. 
$p_z\to p_z+\ab^T$ in case of the external source \eq{infsource}. The spatial spread of such
a local Galilean-boost, shown in the second equation of \eq{homfl}, indicates the presence of 
shear viscosity in the dynamics of the three-current.

One finds similar processes in a collisionless plasma. The retarded solution of the 
collisionless Boltzmann equation in the presence of an electric field and the Maxwell equations produces
Landau damping, the spread of the energy of the electric field over the charges \cite{landau}.
Though there is no collision term in the Boltzmann equation, its simultaneous solution with the
equation of motion for the electric field introduces genuine interactions among the charges.
The spread, described by Eqs. \eq{homfl} corresponds to more elementary processes than 
the Landau damping because it takes place in a truly noninteracting quantum gas. 
A formal similarity between the two phenomena is the dephasing, taking place either
in the solution of the Schr\"odinger equation for ideal quantum gases or in the 
Fourier integral representation of the retarded Lienard-Wiechert potential for 
classical plasma \cite{clemmow}.

One should bear in mind the difference between the spread of the wave packet and
truly diffusive processes. The former is related to the asymmetry of the initial 
and final states in physically motivated bases and arises in time reflection 
symmetric dynamics, governed by the Schr\"odinger equation. The nontrivial, dynamical issue of the time arrow,
the dynamical breakdown of the time reversal invariance appears in the latter 
when irreversibility is observed. The key element is that the soft, gapless 
collective modes have a time scale which is much longer than the characteristic 
time of the observation. The bosonisation is an exact rewriting of the
dynamics of a fermion system with conserved particle number and
it corresponds within the functional formalism to the retaining of all 
Green functions of even order as independent variables when the fermion
fields are integrated over in the path integral expressions. Instead of
such an extremely involved dynamics our calculation can be considered
as an approximative bosonisation where one keeps the density and current,
the first two terms of McLaurin series of the two-point funtion, ignoring
the remaining informations in the two-point funtion and all higher
order Green functions. Such a coarse graining retains the long distance,
low frequency part of the dynamics and the resulting effective theory
applies for scales far in the infrared compared to the intrinsic scale 
of the ideal gas. Such a loss of information related to the strong
separation of scales generates the dynamical breakdown of the time reversal
invariance and irreversibility. Had we included enough informations
in the effective description to extend it to the scale of the ideal gas
we would have lost the dissipative terms.

\section{Summary}
It is pointed out in this paper that the statistics of undistinguishable
and noninteracting particles produces complicated, nonpolynomial interaction`
vertices when the dynamics is diagnosed by composite operators. In the physically
motivated choice of the density and current operators one finds damping and 
decoherence in an ideal gas. These phenomena arise at the only microscopic 
scale of the ideal gas and differ from the analogous phenomena of truly
interactive systems. Namely, the damping leads to irreversibility for processes
far infrared compared to the intrinsic scale of the ideal gas 
and the decoherence of the current operator is stronger at 
shorter length and time scales. These results provide a better background to
interpret damping and the classical limit of interacting systems. In particular,
they clarify the physical content of the UV intial conditions when the
renormalization group method is applied to discover the quantum-classical transition.

\acknowledgments
I thank the number of discussions with Janos Hajdu which gave encouragement and 
insight during the preparation of this work.

\end{document}